# The nanosphere phonon laser

A phonon laser made from a levitated silica nanosphere held in a controllable optical trap offers a useful tool for studying phonon–photon interactions.

Ran Huang and Hui Jing

A phonon laser — a device that performs coherent amplification of phonons, the quanta of vibration — can be considered as an analogue of a conventional laser that performs coherent amplification of photons, the quanta of light. Indeed, phonon lasers share many similarities with lasers: a positive feedback mechanism, a lasing threshold where the system switches from spontaneous to stimulated emission, gain saturation, and linewidth narrowing above the lasing threshold.

An interesting feature of phonon lasers is that their wavelength of operation is considerably shorter than that of a photon laser of the same frequency because vibrations (sound waves) propagate at a much smaller speed than light. This makes phonon lasers and sound waves attractive for non-destructive testing applications such as high-precision imaging and sensing, for example.

Furthermore, phonon lasers offer the potential of on-chip integration with micro- and nanophotonics, creating compact hybrid photonic–phononic systems.

In the last decade, phonon lasers have been built using many different approaches, involving semiconductor superlattices, nanomechanical devices, nanomagnets, single trapped ions, ultracold atomic gas, electromechanical resonators and coupled optical microresonators[1]. Phonon lasers have also been used to study and harness mode competition[2], exceptional point physics[1,3,4] and non-reciprocal devices[5].

Now writing in *Nature Photonics*, Pettit and colleagues introduce a platform for a phonon laser that is based on a levitated neutral particle[6]. The team used optical tweezers to trap a 150-nm-diameter silica glass nanosphere and measured the sphere's mechanical motion by analysing how it scatters a weak coherent light signal (Fig. 1).

To turn this levitated system into a phonon laser, a feedback mechanism is required that is not readily available due to the absence of a cavity. Pettit and colleagues addressed this issue by implementing a measurement-based feedback mechanism that consists of two crucial components, namely a parametric (nonlinear) feedback cooling part and a linear amplification part.

To achieve this feedback, Pettit and colleagues measure the position of the levitated nanosphere in all three degrees of motion using highly sensitive detectors (displacement sensitivity of ~1 pm Hz$^{-1/2}$). The electrical signals generated by the detectors are then split into two parts and used to generate two feedback signals. The first feedback signal is nonlinear in nature and prepared by bandpass filtering, frequency doubling, phase shifting and linear amplification. In contrast, the second feedback signal is linear and prepared using a phase-locked loop (which tracks the phase and frequency of the oscillation), phase shifting and linear amplification of the electrical signal. These two feedback signals are then summed and input to an electro-optic modulator that modulates the power of the trapping laser and hence controls the centre-of-mass motion of the sphere.

The nonlinear feedback helps reduce (cool) the centre-of-mass motion by modulating the trap potential by twice the frequency of the oscillation of the nanosphere. This nonlinear cooling plays the role of gain saturation in an optical laser. However, this is not sufficient to generate mechanical coherence in the motion of the trapped and levitated nanosphere. One needs amplification of the centre-of-mass motion, akin to the optical amplification provided by the gain medium in optical lasers. In this case, this is achieved with the linear feedback signal that includes linear amplification, linear feedback cooling and damping. A threshold occurs and stimulated emission of phonons takes over when amplification overcomes the sum of damping and feedback cooling. Phonon number fluctuations and other types of noise correspond to spontaneous emission.

Levitated optomechanics, where an oscillating object is levitated with a strongly focused laser beam using the forces of light, has attracted much interest in recent years due to the possibilities and benefits it offers. Levitating objects in optical traps helps to isolate them from the environment and hence substantially reduces forces and noise induced by the surroundings (no mechanical clamping). This then opens the door to having control over the motion of objects and even making their motion so quiet that ground-state cooling of a mesoscopic mechanical system is achieved, making it possible to study mechanical motion in the quantum regime. The result is unique tools for testing fundamental theories of quantum physics, such as creating macroscopic quantum superposition states or entanglement generation[7]. An optically levitated dielectric particle can also be used as an ultra-sensitive detector for short-range forces, gravitational waves and dark energy, for example[8]. Recently, rotational degrees of freedom of levitated objects have been studied using a circularly polarized laser, and a nanomechanical rotor with GHz rotation frequency has been realized[9,10].

With the levitated nanosphere and the feedback system developed, Pettit and co-workers have a phonon laser that is easily and precisely controlled. They can tune the oscillation frequency and hence the phonon laser frequency by controlling the d.c. optical power of the trap. They can also vary the modulation of the amplifying feedback to control the pump power for the phonon laser and investigate threshold behaviour, and they can tune the feedback cooling to control the initial phonon number in the centre-of-mass motion and move the threshold value.

A critical feature of a phonon laser is coherence and in order to show that this is present, Pettit and colleagues have performed several measurements.

First, they measured the full phonon probability distribution of their system and discovered Boltzmann statistics below threshold implying a thermal state, and as they pushed the system above threshold the statistics changed to a Gaussian distribution truncated at zero phonon number and a variance greater than the mean. Although the measured phonon distribution far above threshold is not a Poisson distribution (mean phonon number equal to variance), which is a characteristic of a coherent state, it does clearly indicate that the system is not in a thermal state. Since the variance of the distribution is smaller than that of a thermal state with the same mean phonon number, one can say that the system is in a subthermal number squeezed state. With the full phonon distribution at hand, Pettit and co-workers calculated the second-order phonon autocorrelation function at zero time delay, $g^2(0)$, as a function of the phonon laser pump power. This clearly shows that as the system is pushed above the threshold, $g^2(0)$ decreases from the value of 2, which is characteristic of a thermal state, down to 1, which is obtained for a coherent state.

In the second measurement to quantify coherence in their phonon laser, Pettit and colleagues have measured the in-phase and quadrature components of the mechanical motion of the nanosphere with a lock-in amplifier. The result shows a transition of the oscillation of the centre-of-mass motion from a thermal state (Brownian motion), represented by a Gaussian P-function, below the threshold to a coherent state, represented by an annulus in phase space, above the threshold.

These findings are further supported in the transient behaviour of the levitated optomechanical system, which confirm the evolution of the phonon distribution from thermal statistics when the linear gain is switched to a Gaussian distribution after a period of time. They also show that the initial exponential growth of the phonon number saturates due to nonlinear feedback cooling, finally reaching the steady-state value.

Although it remains to be seen whether this levitated nanosphere phonon laser will lead to novel devices, it is certainly a useful new tool for exploring phononics and phonon–photon interactions in new regimes. Ultimately, it may even yield novel applications and better sensors, for example for force and displacement measurement, and imaging systems with considerably improved resolution.

## Authorship and affiliation

Ran Huang and Hui Jing*
*Key Laboratory of Low-Dimensional Quantum Structures and Quantum Control of Ministry of Education, Department of Physics and Synergetic Innovation Center for Quantum Effects and Applications, Hunan Normal University, Changsha 410081, China.*
*e-mail: jinghui73@foxmail.com

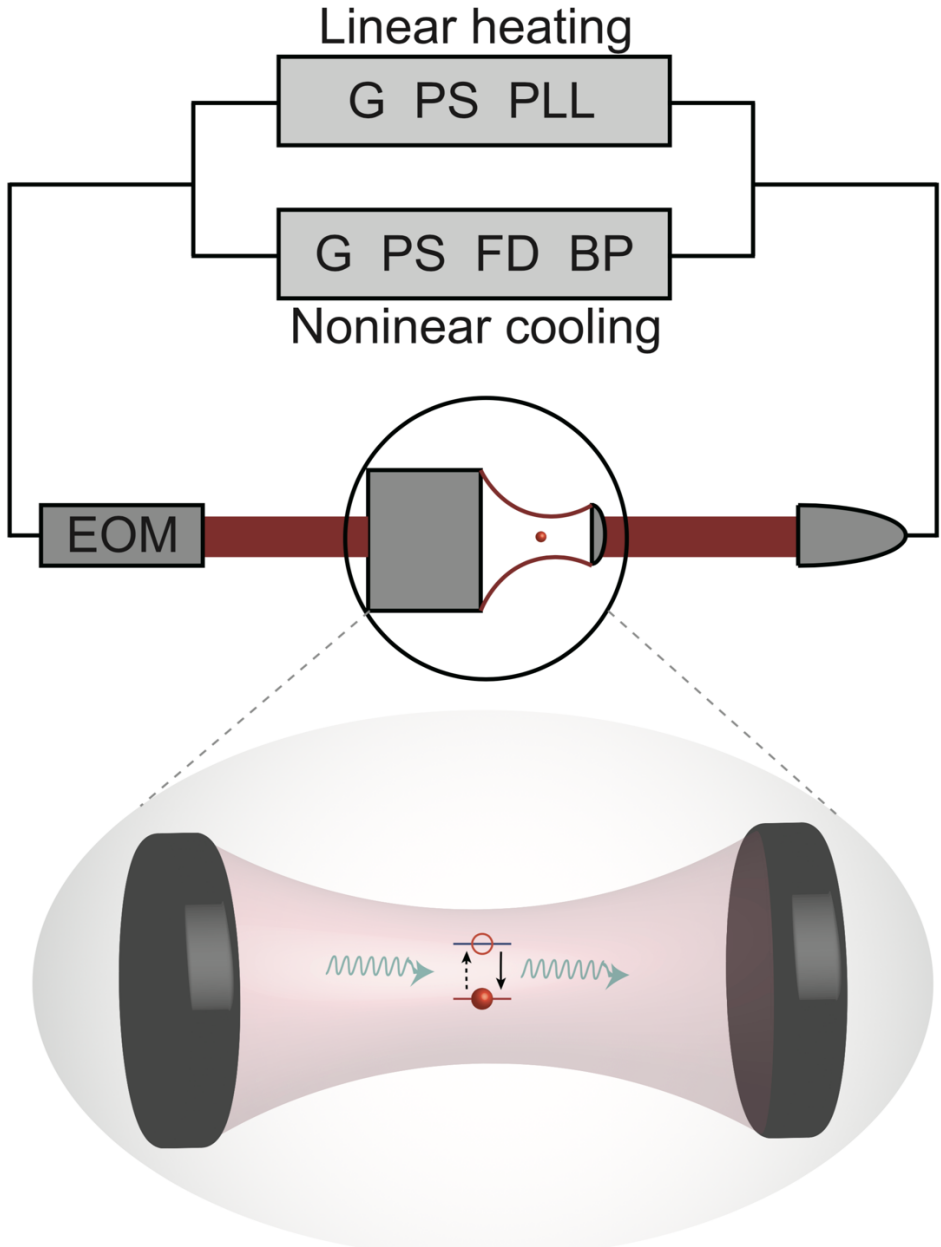

Fig.1 schematic of a phonon laser based on a levitated optomechanical system. A silica nanosphere (red sphere) is levitated in a vacuum by an optical tweezer and its mechanical motion is measured via the scattering of a weak coherent light signal. Phonons (green wavy lines) are either annihilated (dotted black arrow) or created (solid black arrow) by the nanosphere. To turn the system into a laser, nonlinear and linear feedback signals are prepared in two separate pathways and used to control the electro-optic modulator (EOM) that modulates the power of the trapping laser and hence controls the centre-of-mass motion of the sphere. BP, electronic bandpass filter; FD, frequency doubler; PS, phase shifter; G, electronic gain; PLL, phase-locked loop.